\documentclass[12pt]{article}
\usepackage{epsf}

\catcode`@=12
\begin{document}
\thispagestyle{empty}
\begin{flushright} 
UCRHEP-T380\\ 
September 2004\
\end{flushright}
\vspace{0.5in}
\begin{center}
{\LARGE	\bf Polygonal Derivation of\\ the Neutrino Mass Matrix\\}
\vspace{1.5in}
{\bf Ernest Ma\\}
\vspace{0.2in}
{\sl Physics Department, University of California, Riverside, 
California 92521, USA\\}
\vspace{1.5in}
\end{center}

\begin{abstract}\
Representations of the symmetry group $D_n$ of the $n-$sided regular 
polygon have generic multiplication rules if $n$ is prime.  Using $D_n$ 
with $n=5$ or greater, a particular well-known form of the Majorana neutrino 
mass matrix is derived.
\end{abstract}
\vspace{1.2in}

**To appear in Fizika B (Zagreb) memorial issue dedicated to Dubravko Tadic.

\newpage
\baselineskip 24pt

The form of the $3 \times 3$ Majorana neutrino mass matrix ${\cal M}_\nu$ 
has been the topic of theoretical study for some time.  If ${\cal M}_\nu$ 
has less than the full 6 parameters, then there exists at least one 
relationship among masses and mixing angles, which may be tested against 
the increasingly more precise experimental data from neutrino oscillations. 
However, even if such a comparison is successful, the question still remains 
as to why it has such a form.  A possible answer is that it comes from an 
underlying symmetry.  In this paper, it is shown how
\begin{equation}
{\cal M}_\nu^{(e,\mu,\tau)} = \pmatrix{a & c & d \cr c & 0 & b \cr d & b & 0}
\end{equation}
may be derived from $D_n$, the symmetry group of the $n-$sided regular 
polygon, where $n$ is a prime number, equal to or greater than 5.

Consider $D_5$, the symmetry group of the regular pentagon.  It has 10 
elements, 4 equivalence classes, and 4 irreducible 
representations.  Its character table is given by

\begin{table}[htb]
\caption{Character Table of $D_5$.}
\begin{center}
\begin{tabular}{|c|c|c|c|c|c|c|}
\hline
class & $n$ & $h$ & $\chi_1$ & $\chi_2$ & $\chi_3$ & $\chi_4$ \\ 
\hline
$C_1$ & 1 & 1 & 1 & 1 & 2 & 2 \\
$C_2$ & 5 & 2 & 1 & $-1$ & 0 & 0 \\
$C_3$ & 2 & 5 & 1 & 1 & $\phi-1$ & $-\phi$ \\
$C_4$ & 2 & 5 & 1 & 1 & $-\phi$ & $\phi-1$ \\
\hline
\end{tabular}
\end{center}
\end{table}

Here $n$ is the number of elements and $h$ is the order of each element. 
The number $\phi$ is the Golden Ratio (or Divine Proportion) 
known to the ancient Greeks:
\begin{equation}
\phi = {\sqrt 5 + 1 \over 2} \simeq 1.618,
\end{equation}
and satisfies the equation
\begin{equation}
\phi^2 = \phi + 1,
\end{equation}
which implies that
\begin{equation}
\phi^{k+1} = \phi F_{k+1} + F_k,
\end{equation}
where $F_k$ are the Fibonacci numbers.  [Zadar on the Dalmatian coast in 
Croatia is an ancient city with a rich history and a university whose origin 
dates back to 1396.  One person who taught there was Luca Pacioli, whose 
famous work {\it Divina Proportione} (1509) was illustrated by Leonardo 
da Vinci.]

The character of each representation is its trace and must satisfy the 
following two orthogonality conditions:
\begin{eqnarray}
\sum_{C_i} n_i \chi_{ai} \chi^*_{bi} = n \delta_{ab}, ~~~~~ 
\sum_{\chi_a} n_i \chi_{ai} \chi^*_{aj} = n \delta_{ij},
\end{eqnarray}
where $n$ is the total number of elements.  The number of irreducible 
representations must be equal to the number of equivalence classes.

The two irreducible two-dimensional representations of $D_5$ may be chosen 
as follows.  For {\bf 2}, let
\begin{eqnarray}
&& C_1: \pmatrix{1 & 0 \cr 0 & 1}, ~~~ C_2: \pmatrix{0 & \omega^k \cr 
\omega^{5-k} & 0}, ~(k=0,1,2,3,4); \nonumber \\ 
&& C_3: \pmatrix{\omega & 0 \cr 0 & \omega^4}, ~\pmatrix{\omega^4 & 0 \cr 0 
& \omega}, ~~~ C_4: \pmatrix{\omega^2 & 0 \cr 0 & \omega^3}, ~\pmatrix{
\omega^3 & 0 \cr 0 & \omega^2},
\end{eqnarray}
where $\omega = \exp(2 \pi i/5)$, then ${\bf 2'}$ is simply obtained by 
interchanging $C_3$ and $C_4$.  Note that
\begin{equation}
2 \cos (2 \pi/5) = \phi-1,  ~~ 2 \cos (4 \pi/5) = -\phi,
\end{equation}
as expected.

For $D_n$ with $n$ prime, there are $2n$ elements divided into $(n+3)/2$ 
equivalence classes: $C_1$ contains just the identity, $C_2$ has the $n$ 
reflections, $C_k$ from $k = 3$ to $(n+3)/2$ has 2 elements each of order 
$n$.  There are 2 one-dimensional representations and $(n-1)/2$ 
two-dimensional ones. For $D_3=S_3$, the above reduces to the ``complex'' 
representation with $\omega=\exp(2 \pi i/3)$ discussed in a recent review 
\cite{fuji}.

The group multiplication rules of $D_5$ are:
\begin{eqnarray}
&& {\bf 1'} \times {\bf 1'} = {\bf 1}, ~~~ {\bf 1'} \times {\bf 2} = {\bf 2}, 
~~~ {\bf 1'} \times {\bf 2'} = {\bf 2'}, \\
&& {\bf 2} \times {\bf 2} = {\bf 1} + {\bf 1'} + {\bf 2'}, ~~~ 
{\bf 2'} \times {\bf 2'} = {\bf 1} + {\bf 1'} + {\bf 2}, ~~~ 
{\bf 2} \times {\bf 2'} = {\bf 2} + {\bf 2'}.
\end{eqnarray}
In particular, let $(a_1,a_2), (b_1,b_2) \sim {\bf 2}$, then
\begin{equation}
a_1 b_2 + a_2 b_1 \sim {\bf 1}, ~~~  a_1 b_2 - a_2 b_1 \sim {\bf 1'}, ~~~  
(a_1 b_1, a_2 b_2) \sim {\bf 2'}.
\end{equation}
Similarly, in the decomposition of ${\bf 2'} \times {\bf 2'}$, 
$(a'_2 b'_2, a'_1 b'_1) \sim {\bf 2}$, and in the decomposition of 
${\bf 2} \times {\bf 2'}$, $(a_2 a'_1, a_1 a'_2) \sim {\bf 2}$, and 
$(a_2 a'_2, a_1 a'_1) \sim {\bf 2'}$.

The most natural assignment of the 3 lepton families under $D_5$ is
\begin{equation}
(\nu_i,l_i), ~l^c_i \sim {\bf 1} + {\bf 2}.
\end{equation}
Assuming two Higgs doublets $\Phi_1 \sim {\bf 1}$, $\Phi_2 \sim {\bf 1'}$, 
the charged-lepton mass matrix is then of the form
\begin{equation}
{\cal M}_l = \pmatrix{a & 0 & 0 \cr 0 & 0 & b-c \cr 0 & b+c & 0},
\end{equation}
where $a,b$ come from $\langle \phi^0_1 \rangle$, and $c$ from 
$\langle \phi^0_1 \rangle$.  Redefining $l^c_{2,3}$ as $l^c_{3,2}$, 
${\cal M}_l$ becomes diagonal with $m_e = |a|$, $m_\mu = |b-c|$, 
$m_\tau = |b+c|$.

Assuming that neutrino masses are Majorana and that they come from the 
naturally small vacuum expectation values \cite{msm98} of heavy Higgs 
triplets $\xi_1 \sim {\bf 1}$, $\xi_{2,3} \sim {\bf 2}$, then
\begin{equation}
{\cal M}_\nu = \pmatrix{a & c & d \cr c & 0 & b \cr d & b & 0}
\end{equation}
as advertised, where $a,b$ come from $\langle \xi^0_1 \rangle$, and $c = f 
\langle \xi^0_3 \rangle$, $d = f \langle \xi^0_2 \rangle$.  The two texture 
zeros are the result of the absence of a Higgs triplet transforming as 
${\bf 2'}$.  In the case of $D_3=S_3$, there is only one two-dimensional 
representation, hence these zeros cannot be maintained without also making 
$c=d=0$.

The decomposition ${\bf 2} \times {\bf 2} = {\bf 1} + {\bf 1'} + {\bf 2'}$ 
holds not only in $D_5$, but also in $D_n$ with $n$ prime and $n > 5$. 
For example in $D_7$, there are 3 two-dimensional irreducible representations, 
corresponding to the 3 cyclic permutations of
\begin{eqnarray}
C_3 &:& \pmatrix{\omega & 0 \cr 0 & \omega^6}, ~~~ \pmatrix{\omega & 0 \cr 0 
& \omega^6}, \\
C_4 &:& \pmatrix{\omega^2 & 0 \cr 0 & \omega^5}, ~~~ \pmatrix{\omega^5 & 0 
\cr 0 & \omega^2}, \\
C_5 &:& \pmatrix{\omega^3 & 0 \cr 0 & \omega^4}, ~~~ \pmatrix{\omega^4 & 0 
\cr 0 & \omega^3},
\end{eqnarray}
where $\omega = \exp(2 \pi i/7)$.  It is clear that
\begin{equation}
{\bf 2}_1 \times {\bf 2}_1 = {\bf 1} + {\bf 1'} + {\bf 2}_2, ~~~~ 
{\bf 2}_2 \times {\bf 2}_2 = {\bf 1} + {\bf 1'} + {\bf 2}_3,
\end{equation}
etc.  Hence Eq.~(13) is valid in all these symmetries.

Phenomenologically, Eq.~(13) has been studied \cite{study} as an example 
of the class of neutrino mass matrices with two texture zeros.  It was 
first derived from a symmetry ($Q_8$ or $D_4$) only recently \cite{fkmt}.  
Whereas $Q_8$ or $D_4$ allows other forms, $D_n$ with $n$ prime and $n \geq 
5$ allows only Eq.~(13).  Models based on $D_4 \times Z_2$ have also been 
proposed \cite{grimus}.  The 4 parameters of Eq.~(13) imply that 
$m_{1,2,3}$ are related to the mixing angles.  Given the present global 
experimental constraints \cite{mstv}:
\begin{eqnarray}
&& \Delta m^2_{atm} = (1.5 - 3.4) \times 10^{-3}~{\rm eV}^2, ~~~ 
\sin^2 2 \theta_{atm} > 0.92, \\ 
&& \Delta m^2_{sol} = (7.7 - 8.8) \times 10^{-5}~{\rm eV}^2, ~~~
\tan^2 \theta_{sol} = 0.33 - 0.49,
\end{eqnarray}
and $|\sin \theta_{13}| < 0.2$, the allowed region in the $m_3-m_2$ plane 
has been obtained in Ref.~[4].  That figure is reproduced here for the 
convenience of the reader.  It shows that there are lower bounds on 
$m_2$ and $m_3$ and that $m_3 < m_2$ up to about 0.1 eV.  The parameter 
$a$ in Eq.~(13) measures neutrinoless double beta decay and has a lower 
bound of about 0.02 eV in this case.

\begin{figure}[htb]
\begin{center}
\epsfxsize=15cm
\epsfbox{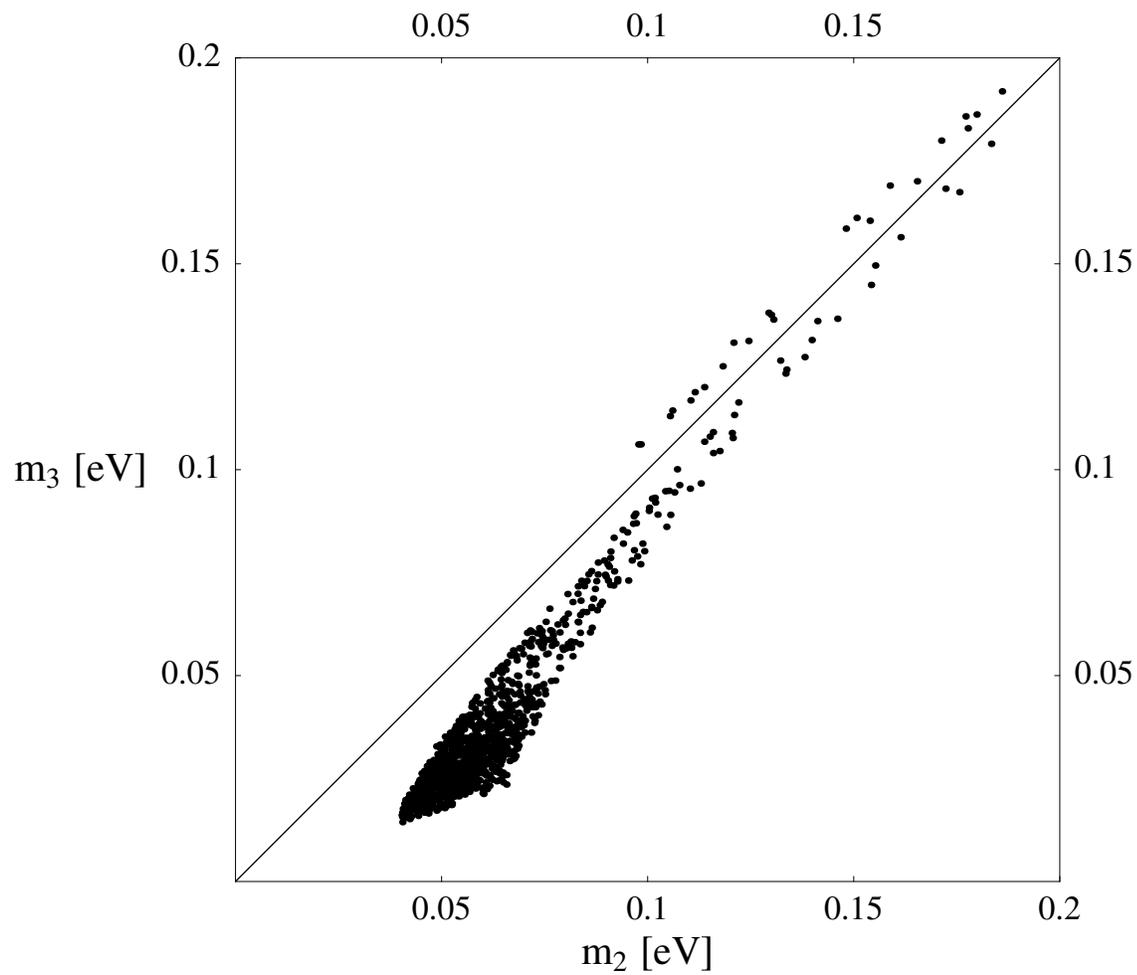}
\caption{Allowed region in $m_2 - m_3$ plane for Eq.~(13)}
\end{center}
\end{figure}

This work was supported in part by the U.~S.~Department of Energy under 
Grant No. DE-FG03-94ER40837.  It is dedicated to the memory of Professor 
Dubravko Tadic.

\bibliographystyle{unsrt}

\begin{thebibliography}{99}

\bibitem{fuji} E. Ma, talk at SI2004, Fuji-Yoshida, Japan, hep-ph/0409075.

\bibitem{msm98} E. Ma and U. Sarkar, Phys. Rev. Lett. {\bf 80}, 5716 (1998); 
E. Ma, Phys. Rev. Lett. {\bf 81}, 1171 (1998).

\bibitem{study} P. H. Frampton, S. L. Glashow, and D. Marfatia, Phys. Lett. 
{\bf B536}, 79 (2002); Z.-Z. Xing, Phys. Lett. {\bf B530}, 159 (2002); {\bf 
539}, 85 (2002); M. Frigerio and A. Yu. Smirnov, Phys. Rev. {\bf D67}, 
013007 (2003).

\bibitem{fkmt} M. Frigerio, S. Kaneko, E. Ma, and M. Tanimoto, hep-ph/0409187.

\bibitem{grimus}
W. Grimus and L. Lavoura,  Phys. Lett. {\bf B572}, 189 (2003); W. Grimus, 
A. S. Joshipura, S. Kaneko, L. Lavoura, and M. Tanimoto, hep-ph/0407112.

\bibitem{mstv} See for example the recent update of M. Maltoni, T. Schwetz, 
M. A. Tortola, and J. W. F. Valle, New J. Phys. {\bf 6}, 122 (2004).

\end{thebibliography}

\end{document}